\documentclass[10pt,a4paper]{article}
\pdfoutput=1
\usepackage[utf8]{inputenc}
\usepackage{xcolor}
\usepackage{fancyhdr}
\usepackage{amsmath,amssymb}
\usepackage{widetext}
\usepackage{multirow,booktabs,graphicx}
\usepackage{mathtools,feynmp-auto}
\usepackage{multirow}
\usepackage{tikz}
\usepackage{pgfplots}
\usepackage{cite}
\usepackage{placeins}
\usepackage{mciteplus}
\usepackage{xspace}
\usepackage[margin=1in]{geometry}
\usepackage[pdfborder={0 0 0}]{hyperref}

\graphicspath{{figures/}}

\newcommand\one{\leavevmode\hbox{\small1\normalsize\kern-.33em1}}

\newcommand{\lag}{\mathcal{L}}

\newcommand{\ope}{\mathcal{O}}

\newcommand{\CP}{C\!P}

\newcommand{\gev}{\text{ GeV}}
\newcommand{\tev}{\text{ TeV}}
\newcommand{\fb}{\text{ fb}}

\newcommand{\pb}{\text{ pb}}

\newcommand{\ifb}{\text{ fb}^{-1}}

\def\slashchar#1{\setbox0=\hbox{$#1$}           
   \dimen0=\wd0                                 
   \setbox1=\hbox{/} \dimen1=\wd1               
   \ifdim\dimen0>\dimen1                        
      \rlap{\hbox to \dimen0{\hfil/\hfil}}      
      #1                                        
   \else                                        
      \rlap{\hbox to \dimen1{\hfil$#1$\hfil}}   
      /                                         
   \fi}

\renewcommand{\Vec}{%
  \mathpalette {\overarrow@\vectfill@}}
\def\vectfill@{\arrowfill@\relbar\relbar{\raisebox{-3.81pt}[\p@][\p@]{$\mathord\mathchar"017E$}}}

\newcommand{\be}{\begin{eqnarray*}}
\newcommand{\ee}{\end{eqnarray*}}

\newcommand{\bee}{\begin{eqnarray}}
\newcommand{\eee}{\end{eqnarray}}
\newcommand{\beeq}{\begin{equation}}
\newcommand{\eeeq}{\end{equation}}

\newcommand{\Sherpa}{S\protect\scalebox{0.8}{HERPA}\xspace}
\newcommand{\Comix}{C\protect\scalebox{0.8}{OMIX}\xspace}
\newcommand{\CSS}{C\protect\scalebox{0.8}{SSHOWER}\xspace}
\newcommand{\Lhapdf}{L\protect\scalebox{0.8}{HAPDF}\xspace}
\newcommand{\Rivet}{R\protect\scalebox{0.8}{IVET}\xspace}

\date{}

\begin{document}

\thispagestyle{empty}
\begin{flushright}
IPPP/20/77 \\
MCnet-21-01
\end{flushright}
\vspace{0.8cm}

\begin{center}

\begin{center}


{\Large\sc Constraining CP violating operators in charged and neutral triple gauge couplings}

\end{center}

\vspace{0.8cm}

\textbf{
Anke Biek\"otter$^{\,a,b}$, 
Parisa Gregg$^{\,a, b}$, 
Frank Krauss$^{\,a, b}$ 
and Marek Sch\"onherr$^{\,a}$}\\

\vspace{1.cm}
{\em {$^a$Institute for Particle Physics Phenomenology, Durham University, United Kingdom}\\
{$^b$Institute for Data Science, Durham University, United Kingdom}
}\\[0.2cm]
\vspace{0.5cm}


\end{center}

\begin{abstract}
We constrain $\CP$-violating charged and neutral anomalous triple gauge couplings using LHC measurements and projections of diboson and VBF $Vjj$ production, both with subsequent leptonic decays. 
For triple gauge couplings involving $W$~bosons we analyse differential asymmetries and interpret our results in the SMEFT at dimension-six. For neutral triple gauge couplings, which are dominantly constrained by high transverse-momentum bins, we present the resulting bounds in terms of a general anomalous couplings framework.
\end{abstract}


\section{Introduction}

The observation of the Higgs boson in 2012~\cite{Aad:2012tfa,Chatrchyan:2012xdj} has been a milestone in the confirmation of electroweak symmetry breaking (EWSB). 
Since, apart from fixing the Higgs couplings, the mechanism of EWSB also predicts the interactions of the electroweak gauge bosons,
precise measurements of the triple gauge couplings (TGCs) play a crucial role in experimentally testing the SM. 
$\CP$-violating interactions of the gauge bosons are of particular relevance in this regard, since they provide additional sources of $\CP$ violation, necessary to describe, for example, electroweak baryogenesis~\cite{Sakharov:1967dj,Kuzmin:1985mm,Shaposhnikov:1987tw,Nelson:1991ab,Morrissey:2012db}. 

In our work, we study $\CP$-odd anomalous triple gauge couplings involving two (charged) $W$~bosons, $WWZ$ and $WW\gamma$, as well as interactions of neutral gauge bosons, $ZZZ$, $ZZ\gamma$ and $Z \gamma \gamma$, which are completely absent in the SM. 
For charged anomalous triple gauge couplings, we consider constraints from the measurement of $WW \to \ell \ell' \nu \nu$~\cite{Aaboud:2019nkz}, $WZ \to \ell^+ \ell^- \ell^\pm \nu$~\cite{Aaboud:2018ohp}, $W\gamma \to \ell^\pm \nu \gamma$~\cite{CMS:2020olm}, $Zjj \to \ell^+ \ell^- j j$~\cite{Aad:2020sle} and $Wjj \to \ell \nu jj$ production~\cite{Sirunyan:2019dyi}.
To describe small deviations from the Standard Model (SM) values of the charged TGCs in a model-independent fashion, we will use the language of Standard Model Effective Field Theory (SMEFT)~\cite{Buchmuller:1985jz, Georgi:1994qn, Grzadkowski:2010es, Alonso:2013hga,Brivio:2017vri},
where $\CP$-odd SMEFT operators influencing the charged TGCs appear at dimension six. 
Constraints on these operators have been studied and constrained in Higgs boson~\cite{Ferreira:2016jea,Brehmer:2017lrt,Bernlochner:2018opw,Englert:2019xhk,Cirigliano:2019vfc,Biekotter:2020flu} and diboson production processes~\cite{Kumar:2008ng,Dawson:2013owa,Gavela:2014vra,Azatov:2019xxn} as well as vector boson scattering~\cite{Ethier:2021ydt}. 
Recently, $\CP$ violation in diboson production has also been studied in Ref.~\cite{DasBakshi:2020ejz}.
We consider the same experimental inputs, but our analysis differs in the selection of observables sensitive to $\CP$ violation, using differential asymmetries rather than complete differential distributions, reducing both experimental and theoretical systematic uncertainties. Our study provides an independent confirmation of the results found in Ref.~\cite{DasBakshi:2020ejz} using \Sherpa for the generation of both the SM as well as the beyond SM events, and they also serve as independent validation of the implementation of this sector in \Sherpa.

For neutral anomalous triple gauge couplings we consider constraints from the $ZZ \to 4 \ell$ and $2\ell 2\nu$ final states~\cite{Aaboud:2017rwm, Aaboud:2019lgy} as well as $Z\gamma \to 2\ell \gamma$ and $2\nu \gamma$ production channels~\cite{Aad:2019gpq,Aaboud:2018jst}. 
Due to the dominance of squared neutral triple gauge coupling (nTGC) contributions compared to the polarization-suppressed (linear) interference with the SM~\cite{Baur:1992cd,Dawson:2013owa,Rahaman:2018ujg}, we investigate their effects on the cross section in the high-$p_T$ regime, rather than studying asymmetries.

\section{Charged aTGC}
\label{sec:charged_aTGC}

SMEFT~\cite{Buchmuller:1985jz, Georgi:1994qn, Grzadkowski:2010es, Alonso:2013hga,Brivio:2017vri} provides a versatile framework to describe small deviations from the SM, such as those induced by anomalous triple gauge couplings. 
In the Warsaw basis~\cite{Grzadkowski:2010es}, there are two dimension-six operators leading to $\CP$~violation in diboson production through a modification of the $\gamma WW$ and $ZWW$ interactions.  We can describe them through the effective Lagrangian
\begin{equation}
    \lag = \lag_\text{SM} 
        + \frac{c_{\tilde{W}}}{\Lambda^2} \,  \ope_{\tilde{W}}
        + \frac{c_{H\tilde{W}B}}{\Lambda^2} \,  \ope_{H\tilde{W}B} \, ,
\end{equation}
where $\lag_\text{SM}$ is the SM Lagrangian, $\Lambda$ denotes the new physics scale and the $c_i$ are the Wilson coefficients of the operators 
\begin{equation}
    \ope_{\tilde{W}} = \epsilon^{IJK} \, \tilde{W}_\mu^{I \nu} \, W_\nu^{J \rho} \, W_\rho^{K \mu} \, ,
    \,
    \ope_{H\tilde{W}B} = H^\dagger \tau^I H \, \tilde{W}^I_{\mu \nu} \, B^{\mu \nu} \, .
\end{equation}
The operator $\ope_{H\tilde{W}B}$ also affects Higgs-gauge couplings and its Wilson coefficient can thus be constrained independently through Higgs sector observables.

\begin{figure}[thb]
    \centering
    \includegraphics[width=.46\textwidth]{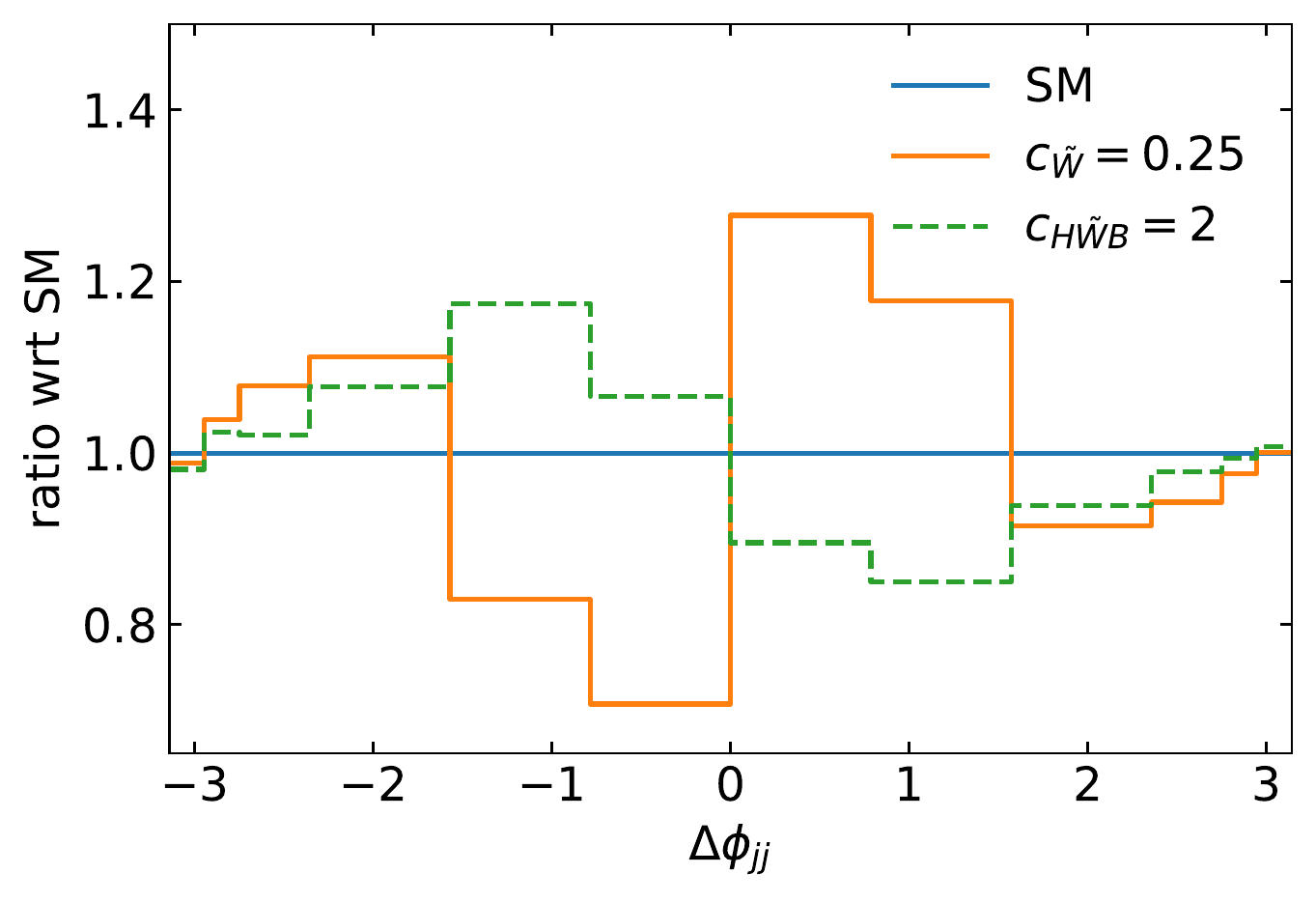}
    \caption{$\Delta \phi_{jj}$ distribution for the $Zjj$ analysis.}
    \label{fig:Zjj_deltaphi}
\end{figure}

In the following, we calculate and combine the constraints on the Wilson coefficients $c_{H\tilde{W}B}$ and $c_{\tilde{W}}$ from $WW$, $WZ$, $W\gamma$ and VBF $Zjj$ and $Wjj$ production at the LHC.
For each considered channel~$i$, we study the differential distributions of an  angle $\zeta_i$ which is defined from the triple products or, equivalently, the difference in azimuthal angle of the rapidity-ordered final-state (pseudo-)particles $k$ and $l$, e.g.~$ \Delta \phi_{kl}\propto \sin^{-1}((\vec{p}_k - \vec{p}_l)_z \, (\vec{p}_k \times \vec{p}_l)_z)$.
The operators $\ope_{\tilde{W}}$ and $\ope_{H\tilde{W}B}$ produce modulations in these distributions. 
As an example, we display the $\Delta \phi_{jj}$ distribution for the two tagging jets in $Zjj$~production in Fig.~\ref{fig:Zjj_deltaphi}. 
We divide each of these $\Delta \phi$ distributions into six pairs of bins using the boundaries
$\pm[0, \, \tfrac{1}{4}, \, \tfrac{1}{2}, \, \tfrac{3}{4},\, \tfrac{7}{8},\, \tfrac{15}{16},\, 1]\cdot\pi$ and construct six asymmetries~$A_{ij}$ between the number $N$ of events in corresponding bins at positive or negative $\Delta\phi$, 
i.e.~we compare the number of events in the bin $[-\tfrac{\pi}{4}, \, 0]$ with the number of events in the bin $[0, \, \tfrac{\pi}{4}]$ and so on:
\begin{equation}
    A_{ij} = \frac{N_{i, -j} - N_{i, +j}}{N_{i, -j} + N_{i, +j}} \qquad i = WW, \, WZ , \, W\gamma , \, Zjj, \, Wjj, \quad 
    j=1, \, \dots ,  \, 6  \, .
\end{equation}
In the above, $N_{i, j}$ corresponds to number of events in bin~$j$ of the $\Delta\phi$ distribution for channel~$i$.
The resulting differential asymmetries $A_{ij}$ are defined such that they vanish exactly for the SM and SM backgrounds, where no $\CP$ violation is present.\footnote{
  The $\CP$ violation present in the SM have been checked to be negligible for the range of observables and coefficients considered in this letter.
} The study of asymmetries has the advantage that systematic uncertainties largely cancel in the ratio and the limits are therefore entirely determined by the statistical uncertainties. 
For channels with a signal-over-background ratio smaller than one, we take into account the uncertainty from background subtraction as 
$\sigma_{A_{ij}}^\text{bkg}=\sqrt{N_{ij}^\text{bkg}}/N_{ij}^\text{sig}$.

We generate events at leading order (LO) with \Sherpa-2.2.10~\cite{Gleisberg:2008ta,Bothmann:2019yzt} with the default \texttt{NNPDF30\_nnlo\_as\_0118} parton distribution function~\cite{Ball:2014uwa} from \Lhapdf 6.2.1~\cite{Buckley:2014ana}; matrix elements are calculated with \Comix~\cite{Gleisberg:2008fv} and parton showered with \CSS~\cite{Schumann:2007mg}. QED corrections are effected through a YFS soft-photon resummation \cite{Yennie:1961ad,Schonherr:2008av}. For multi-parton interactions, hadronisation, and subsequent hadron decays we use the \Sherpa default settings. EFT contributions are generated using the \texttt{SMEFTsim} model~\cite{SMEFTsim} in \Sherpa through its UFO \cite{Degrande:2011ua} interface \cite{Hoche:2014kca}. We consider the interference of the SM with the dimension-six operator only and neglect contributions from the squares of dimension-six terms.

In each channel, we normalize the SM cross section to the experimentally observed cross section and assume identical normalization factors for the SM and the EFT contributions. 
To take into account detector effects, we include a flat detector efficiency which we deduce from the ratio of the predicted cross section and the predicted number of events provided by the experimental collaborations
$ \epsilon_\text{det} = N_\text{events, pred}/(\sigma_\text{pred} \, \lag_\text{int})$.

\paragraph{$WW$ production.}
For $WW$ production, we consider an asymmetry in the sine of the difference of the azimuthal angles~$\phi$ of the two final state leptons ordered by their pseudorapidity,~$\zeta_{WW} = \Delta \phi_{\ell \ell}$.
We make use of the existing \Rivet~\cite{Buckley:2010ar} analysis to reproduce the experimental cuts and normalize the \Sherpa cross section to the measured value of $\sigma_{\text{fid}, EW} = 379.1 \pm 27.1 \fb$~\cite{Aaboud:2019nkz}. 
The detector efficiency is deduced from the difference between the predicted cross section 
and the predicted number of events 
$\epsilon_\text{det} = 0.61$.
Since the signal-over-background ratio $S/B > 1$, we can safely neglect the uncertainty from background subtraction.

\paragraph{$WZ$ production.}
For $WZ$ production, the $\CP$-sensitive observable considered is~$\zeta_{WZ} = \Delta \phi_{Z \ell'}$, 
where $\ell'$ denotes the lepton from the decay of the $W$~boson and $Z$ denotes the reconstructed $Z$~boson from the same-\-flavor-\-opposite-\-sign lepton pair. 
We normalize the \Sherpa cross section to the measured value of $\sigma_{\text{fid}, EW} = 254.7 \pm 11.5 \fb$~\cite{Aaboud:2018ohp} and assume a detector efficiency of $\epsilon_\text{det} = 0.52$.
Since the signal-over-background ratio $S/B > 1$, we can safely neglect the background contributions.

\paragraph{$W\gamma$ production}
For $W\gamma$ production in the $\ell \nu \gamma$ final state we define the $\CP$~sensitive observable~$\zeta_{W\gamma} = \Delta \phi_{\gamma \ell}$, 
where $\ell$ and $\gamma$ denote the lepton from the $W$~boson decay and of the photon, respectively. 
CMS has performed an analysis for $W\gamma$ production at $13$~TeV for an integrated luminosity of $\lag = 127.1 \ifb$~\cite{CMS:2020olm}. 
Including the decay of the $W$~boson, the analysis has measured a cross section of $\sigma_\text{fid} = (3.32 \pm 0.16) \pb$. 
We implemented the experimental cuts in \Rivet and normalized the cross section after cuts to this value. 
From the expected number of signal events and the expected cross section, we deduce a detector efficiency of $\epsilon_\text{det} = 0.59$. 
For $W\gamma$ production, the signal-over-background ratio after cuts is $S/B \approx 0.54$ and we therefore explicitly take the uncertainty from background subtraction into consideration. Since the dominant background contributions arise from experimental effects such as nonprompt leptons and photons and $e$-induced photons, it is difficult to estimate their shape. For this reason, we assume the background shape to closely follow the signal shape in $\Delta \phi_{\gamma \ell}$.

\paragraph{$Zjj$ production}
In vector boson fusion $Zjj$ production, $\CP$ violation in the $ZWW$ and $\gamma WW$ couplings causes modulations in the $\Delta \phi_{jj}$~distribution of the $\eta$-ordered jets, see Fig.~\ref{fig:Zjj_deltaphi}.
We normalize the \Sherpa cross section to the measured value of $\sigma_{\text{fid}, EW} = 37.4 \pm 6.5 \fb$~\cite{Aad:2020sle} and take into account a factor of $\epsilon_\text{det} = 0.85$ for detector effects. 
Since $S/B \approx 0.59$, we consider the uncertainty from background subtraction using the  $\Delta \phi_{jj}$ distribution for the background as given in the experimental reference. 

\paragraph{$Wjj$ production}
For VBF $Wjj$ production, we again base our analysis on the $\Delta \phi_{jj}$~distribution of the $\eta$-ordered jets. 
On top of the baseline selection used in Ref.~\cite{Sirunyan:2019dyi}, we apply a stricter cut on the invariant mass of the tagging jets $m_{jj} > 1100 \gev$, resulting in a signal-over-background ratio of $S/B \approx 0.13$.
For the background, we have generated the dominant QCD $Wjj$ contribution with \Sherpa to obtain the shape. We normalize the event numbers to match the predicted number of total signal and background events in Ref.~\cite{Sirunyan:2019dyi} rescaled by the luminosity. 

\paragraph{Combination.}
We combine the constraints on the Wilson coefficients from measurements of the $WW$, $WZ$, $W\gamma$, $Zjj$ and $Wjj$ channels in a $\chi^2$ analysis. 
Since systematic uncertainties cancel out in our observables, we do not need to consider correlations between uncertainties of the different channels and directly calculate the $\chi^2$ from the differential asymmetries $A_{ij}$ via 
\begin{equation}
    \chi^2 = \sum_{i, j} \frac{(A_{ij}- 0.)^2}{\sigma_{A_{ij}}^2} \, , 
    \qquad 
    i = WW, \, WZ, \, W\gamma, \, Zjj ,\, Wjj,
    \quad 
    j=1, \, \dots ,  \, 6  \, ,
\end{equation}
where $\sigma_{A_{ij}}$ denotes the combined statistical uncertainty from signal and background on the asymmetry in bin~$j$ of channel~$i$. 

We present the expected results for LHC Run~II with an integrated luminosity of $\lag_\text{int} = 139 \ifb$ as well as prospects for the high luminosity LHC with an integrated luminosity $\lag_\text{int} = 3000 \ifb$ in Fig.~\ref{fig:combi_aTGC}. 
The strongest constraints result from $W\gamma$ production for $c_{H\tilde{W}B}$ and from the $Zjj$ and $Wjj$ channels for $c_{\tilde{W}}$.
Our bounds approximately agree with those presented in Ref.~\cite{DasBakshi:2020ejz}\footnote{Notice that our paper has a sign difference for the operator $c_{H\tilde{W}B}$ with respect to Ref.~\cite{DasBakshi:2020ejz} and Ref.~\cite{Aad:2020sle}. 
We have validated our results by detailed comparison with MadGraph, with identical results. 
The one-parameter limits presented are not affected by the sign change.}. 
Some differences occur due to the inclusion of detector inefficiencies. 
For the $W\gamma$ channel, we benefit from being able to recast an existing 13 TeV analysis rather than relying on assumptions for the cuts. 
Therefore, the cross section used for this channel is a factor $10$ smaller in our analysis than assumed in Ref.~\cite{DasBakshi:2020ejz}.

Our limit on $c_{H\tilde{W}B}$ is much stronger than the bound resulting from Higgs observables. 
At $3000 \ifb$ luminosity the expected limits are 
$|c_{H\tilde{W}B}|/\Lambda^ 2 < 3.1 \tev^{-2}$ from Higgs WBF+$\gamma$ production~\cite{Biekotter:2020flu} and 
$|c_{H\tilde{W}B}|/\Lambda^ 2 < 1.5 \tev^{-2}$ from standard Higgs production processes~\cite{Bernlochner:2018opw} respectively compared to  $|c_{H\tilde{W}B}|/\Lambda^ 2 < 0.04 \tev^{-2}$ for this analysis of diboson observables. 
A fit combination of Higgs and diboson observables could in turn further improve the limits on other Wilson coefficients currently constrained from Higgs observables, $c_{H\tilde{G}}$, $c_{H\tilde{W}}$ and $c_{H\tilde{B}}$. 
The Wilson coefficient of the operator $\ope_{\tilde{W}}$ is constrained to $|c_{\tilde{W}}|/\Lambda^ 2 < 0.02 \tev^{-2}$ in our fit. 
High-luminosity LHC projections for diboson plus vector boson scattering data using distributions up to high-$p_T$ instead of actual $\CP$-sensitive observables find competitive constraints~\cite{Ethier:2021ydt}, further highlighting the necessity to combine fits of all available LHC data sets.

\begin{figure}[thb]
    \centering
    \includegraphics[width=.46\textwidth]{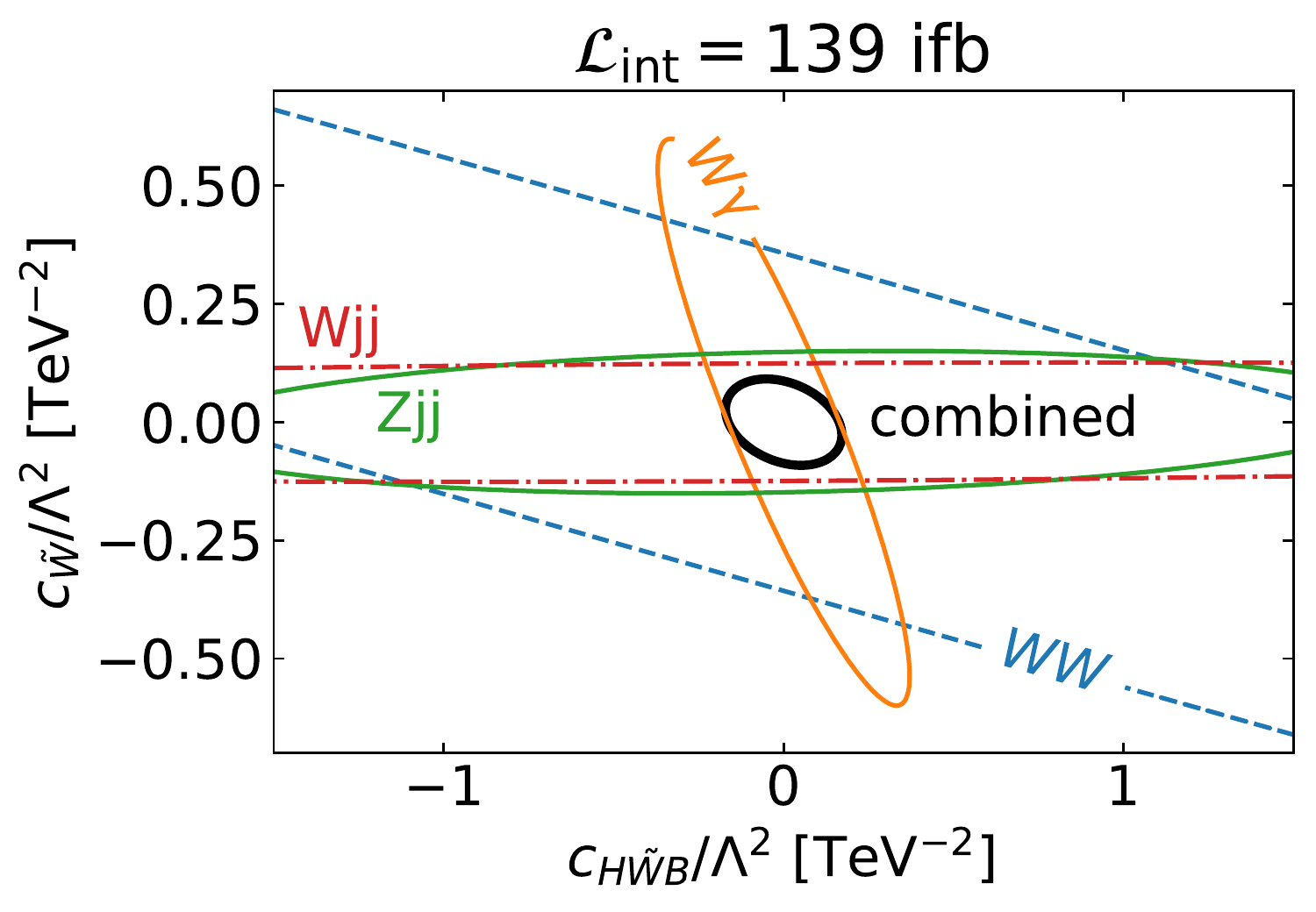}
    \quad
    \includegraphics[width=.46\textwidth]{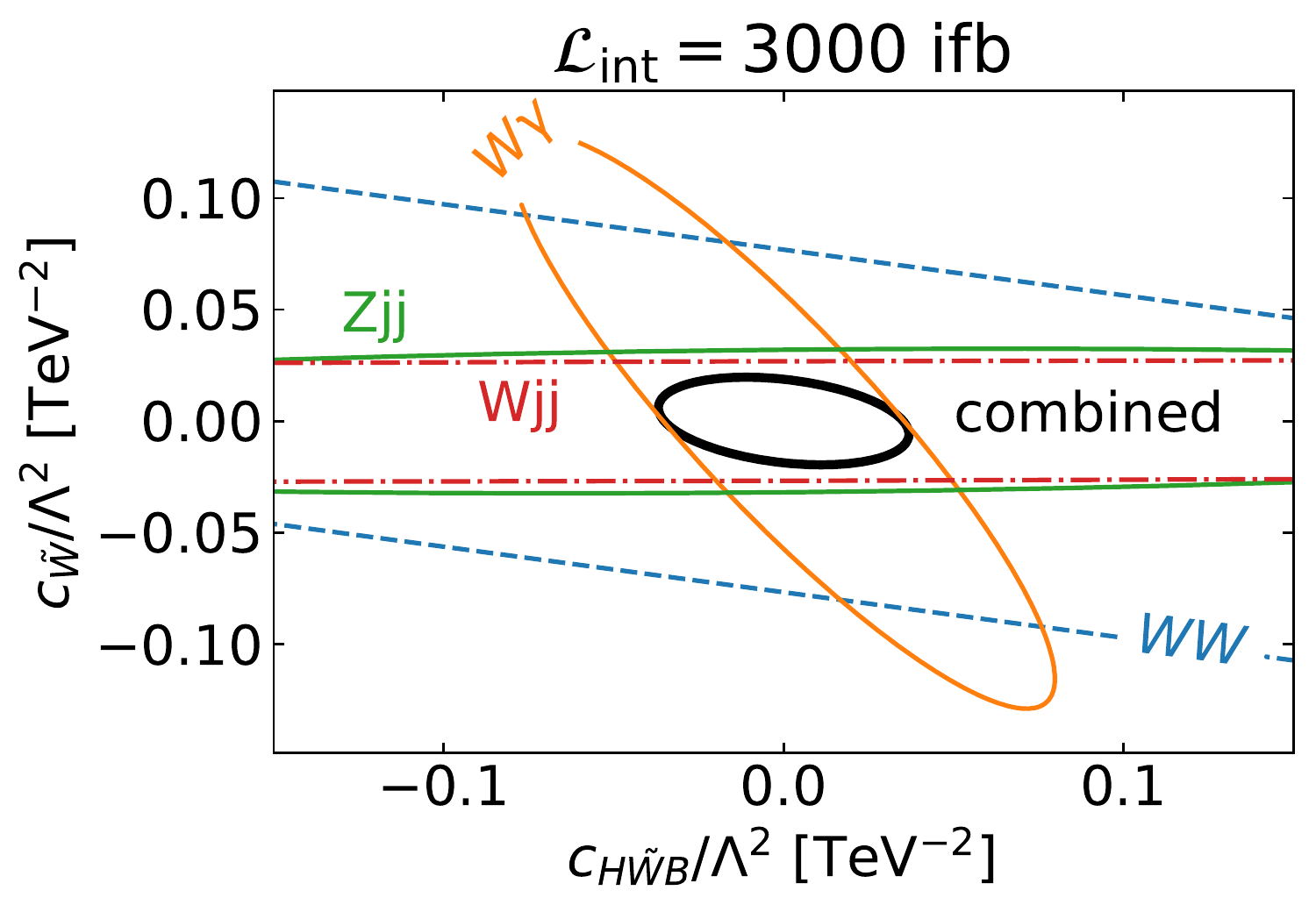}
    \caption{Combination of the $95\%$~CL limits from $WW$, $WZ$, $W\gamma$ and $Zjj$ production after LHC Run~II ($\lag_\text{int} = 139 \ifb$,  left) and after the HL-LHC ($\lag_\text{int} = 3000 \ifb$, right). The limits from $WZ$ production are too weak to be visible in the plots. }
    \label{fig:combi_aTGC}
\end{figure}

\FloatBarrier
\section{Neutral aTGCs}
\label{sec:neutral_aTGC}
Neutral triple gauge couplings are absent in the SM at LO. 
Therefore, the observation of these couplings would be a clear hint for physics beyond the SM~\cite{Corbett:2017ecn}.
The most general parametrization of nTGCs in $ZZ$ and $Z\gamma$ production is given by \cite{Gounaris:1999kf}, cf.\ also~\cite{Hagiwara:1986vm,Baur:1992cd,Baur:2000cx}, 
\begin{equation}
  \begin{split}
  \lag
  = \lag_\text{SM}
    + \frac{e}{M_Z^2}
      \bigg[
      & - [f_4^\gamma \, (\partial_\mu F^{\mu \beta}) + f_4^Z \, (\partial_\mu  Z^{\mu\beta})]
          Z_\alpha (\partial^\alpha Z_\beta)
        + [f_5^\gamma \, (\partial^\sigma F_{\sigma\mu}) + f_5^Z \, (\partial^\sigma  Z_{\sigma\mu} )]
          \widetilde{Z}^{\mu\beta} Z_\beta \\
      & - [h_1^\gamma \, (\partial^\sigma F_{\sigma \mu}) + h_1^Z \,(\partial^\sigma  Z_{\sigma\mu})]
          Z_\beta F^{\mu \beta}
        - [h_3^\gamma \, (\partial_\sigma F^{\sigma\rho}) + h_3^Z \,(\partial_\sigma  Z^{\sigma\rho} )]
          Z^\alpha\widetilde{F}_{\rho\alpha}  \\
      & - \frac{1}{M_Z^2}
          \Big\{  h_2^\gamma \,[\partial_\alpha \partial_\beta \partial^\rho F_{\rho\mu}]
                + h_2^Z\, [\partial_\alpha \partial_\beta (\square + M_Z^2) Z_\mu]
          \Big\} Z^\alpha F^{\mu \beta} \\
      & + \frac{1}{2 M_Z^2}
          \Big\{  h_4^\gamma \, [\square\,\partial^\sigma F^{\rho\alpha}]
                + h_4^Z \, [ (\square + M_Z^2) \partial^\sigma Z^{\rho \alpha}]
          \Big\} Z_\sigma \widetilde{F}_{\rho\alpha}
      \bigg] \; .
  \end{split}
  \label{eq:nTGC_para}
\end{equation}
Non-zero coefficients $f_4^V$, $h_1^V$ and $h_2^V$ lead to $\CP$-violating interactions, while the coefficients $f_5^V$, $h_3^V$ and $h_4^V$ parametrize $\CP$-conserving $ZZZ$, $ZZ\gamma$ and $Z\gamma\gamma$ interactions.
In the SM, all $h_i$ and $f_i$ are zero at tree level. 
At the one-loop level, however, the $\CP$-conserving couplings $f_5$, $h_3$ and $h_4$, receive non-zero contributions with relative sizes at the order of $\ope(10^{-4})$~\cite{Gounaris:2000tb}.

In contrast to charged TGCs, neutral TGCs can currently only be constrained to a regime where the quadratic terms clearly dominate over the linear interference terms with the SM, which are suppressed by the allowed polarizations of the gauge bosons. 
As a result, bounds on $\CP$-violating neutral triple gauge couplings (nTGCs) stem primarily from their effect on the cross section in the high-$p_T$ regime and they do not come from $\CP$-sensitive observables~\cite{Baur:1992cd,Dawson:2013owa,Rahaman:2018ujg}.
In this section, we will therefore study the enhancement of relevant cross sections in high-momentum bins of kinematic distributions rather than $\CP$ asymmetries. 
In particular, we will study the high-$p_T$ regime of the distributions of $p_T^{\ell \ell}$ in $ZZ$ production as well as $E_{T,\gamma}$ in $Z\gamma$ production. 
Bounds on nTGCs have previously been discussed for the LHC~\cite{Baur:1992cd,Dawson:2013owa,Degrande:2013kka,Rahaman:2018ujg} as well as for future lepton~\cite{Rahaman:2016pqj,Ellis:2019zex} 
and proton colliders~\cite{Yilmaz:2019cue, Senol:2019swu, Ellis:2020ljj}.
Since both the bounds on the coefficients of $\CP$-violating interactions and their $\CP$-conserving counterparts ($f_4^V \leftrightarrow f_5^V$, $h_1^V \leftrightarrow h_3^V$, $h_2^V \leftrightarrow h_4^V$) result from their quadratic effect on the cross section in the high-$p_T$ regime, their limits are typically very similar. 

Neutral triple gauge couplings do not appear at the dimension-six level in the SMEFT. 
They are, however, induced at dimension-eight~\cite{Degrande:2013kka} ($f_4^V$, $f_5^\gamma$, $h_1^V$, $h_3^Z$) or even higher dimension. 
While an interpretation of nTGCs in  SMEFT at dimension-eight is therefore possible, the clear dominance of the quadratic terms over the dimension-eight interference terms renders the interpretation cumbersome and possibly flawed. 
Consequently, we will rely on the parametrization given in Eq.~\eqref{eq:nTGC_para}.

Events for the analysis of neutral anomalous gauge couplings are generated at leading order using the native \texttt{SM+AGC} model in \Sherpa-2.1.1 \cite{Gleisberg:2008ta} as well as an implementation in a UFO model \cite{Bothmann:2019yzt,Hoche:2014kca,ShankhaUFONTGC}.
Event generation includes both the suppressed and mostly negligible interference with the SM model as well as the squared nTGC contributions.

\paragraph{$ZZ$ production.}
We study $ZZ$~production in its leptonic $4\ell$~\cite{Aaboud:2017rwm} and $2\ell 2\nu$~\cite{Aaboud:2019lgy} final states. 
The measured cross sections in the fiducial regions of these channels are $\sigma_{4\ell} = (46.2 \pm 2.4) \fb$~\cite{Aaboud:2017rwm} and $\sigma_{2\ell2\nu} = (25.4 \pm 1.7) \fb$~\cite{Aaboud:2019lgy}, respectively.
In both cases, we use the $p_T^{\ell \ell}$ distributions to constrain the nTGC; in the $4\ell$ final state we use the two leptons of the leading reconstructed $Z$~boson. 
To facilitate direct comparison with published data, we employ the binning used by the experimental collaborations for their luminosity projections.

In our event generation, we include the LO $gg$ and $qq$ initial state contributions for $ZZ$ production. The effect of nTGCs is, however, only included for the $qq$ intial state which makes up for about $90 \%$ of the total number of events. 
NNLO QCD and NLO EW corrections for the events are included through bin-by-bin $k$~factors, assuming the same values for SM and BSM contributions.
These are deduced from the ratio of the LO results with respect to the most precise \Sherpa prediction available. The total number of events in each bin $i$ is given by $N_i^\text{\Sherpa} = N_i^{qq} + N_i^{gg} = \epsilon_i^\text{det} \, \lag_\text{int} \, (\sigma_i^{qq, \text{NLO}} + 1.67 \, \sigma_i^{gg})$, where the $gg$ contribution is corrected by a relative $k$~factor of $1.67$. 
Detector effects are accounted for through bin-by-bin detector efficiency factors~$\epsilon_i^\text{det}$ for the $4\ell$ final state (ranging between $0.57$ and $0.69$) while we use a global detector efficiency of $\epsilon^\text{det} = 0.57$ for the $2\ell 2 \nu$ analysis. 
\medskip

\begin{table}[thb]
    \centering
    \begin{tabular}{lcccc}
    \toprule
    luminosity [$\ifb$]     &  $|f_4^\gamma| \times 10^4$ & $|f_4^Z| \times 10^4$\\
         \midrule
    $139$  & $11.$ & $9.1$ \\
    $300$  & $9.1$ & $7.7$ \\
    $3000$  & $7.2$ & $6.1$  \\
    \midrule
    $300$ (half syst)  & $8.2$ & $7.0$ \\
    $3000$ (half syst)  & $5.3$ & $4.5$ \\
    \bottomrule
    \end{tabular}
    \caption{Expected limits on nTGCs for the combination of the $ZZ \to 4 \ell$ and $ZZ \to 2 \ell 2 
    \nu$ analyses at different luminosities. The limits on the parameters $f_5^V$ which lead to $\CP$-conserving interactions are equivalent to those on their $\CP$-violating counterparts.
     In the two bottom rows, we present the limits assuming that the relative systematic uncertainties in each bin have been halved with respect to the value quoted by the experimental collaborations at $36.1\ifb$.}
    \label{tab:limits_ZZ_ana}
\end{table} 
To set limits on the nTGCs, we perform a $\chi^2$ analysis for each bin in the two available $p_T^{\ell \ell}$ distributions,
\begin{equation}
   \chi^2 =  \sum_{i \in \text{bins}} \frac{(N_i^\text{data}-N_i^\text{pred})^2}{N_i^\text{data}+(\sigma_i^\text{syst})^2} \, ,
   \label{eq:chi2_nTGC}
\end{equation}
where $N_i^\text{data}$ and $N_i^\text{pred}$ denote the number of observed and predicted events in each bin and $\sigma_i^\text{syst}$ is their systematic uncertainty. 
For both analysis channels, the constraints on nTGCs stem almost entirely from the last bin, i.e.~$p_T^{\ell \ell} \in [555, \,  3000]\gev$ 
in $4\ell$ final state and 
$p_T^{\ell \ell} \in [350, \, 1000] \gev$ in the $2\ell 2\nu$ final state.

To validate our analysis, we have explicitly checked that we can reproduce the limits on $f_i^V$ presented by the experimental collaborations~\cite{Aaboud:2017rwm,Aaboud:2019lgy} for a luminosity of $36.1 \ifb$ at the $15 \%$ level. 
Deviations from those limits can be fully explained by the use of different Monte Carlo generators and the fact that we only know the global detector acceptance rather than a bin-by-bin value for the $2\ell 2\nu$ final state.

Combining the limits from the $4\ell$ and $2 \ell 2 \nu$ final states for a luminosity of $3000\ifb$, we find $95\,\%$~CL bounds of 
\begin{equation}
    |f_4^\gamma| < 7.2 \times  10^{-4} \, , \quad 
    |f_4^Z| < 6.1 \times 10^{-4} \, , 
\end{equation}
for the parameters inducing $\CP$-violating interactions. Since the linear interference contributions are not statistically relevant, we display the limit on the absolute values of the parameters instead of presenting separated upper and lower $95\,\%$~CL limits. 
We collect projected limits at different luminosities in Tab.~\ref{tab:limits_ZZ_ana}. 
As expected, the limits on the parameters $f_5^V$ which lead to $\CP$-conserving interactions are equivalent to those of its $\CP$-violating counterparts~$f_4^V$.
Our combined $139\ifb$ limits approximately agree with those found by CMS for LHC Run-II in the $4\ell$~final state~\cite{Sirunyan:2020pub}, which however draws most of its sensitivity from an overflow bin, $m_{ZZ} > 1300 \, \gev$.

Had we included the overflow in the last bin instead of keeping to the binning used by the experimental collaborations, the obtained limits would have tightened by $\lesssim 20\,\%$. 
We generally avoid including the overflow in our last bins, however, to make sure that all considered events lie in a kinematic regime for which the detector is well understood. 
In addition, using a constrained last bins circumvents potential issues when translating the limits to other frameworks such as EFTs. 

\paragraph{$Z\gamma$ production.}
We study $Z\gamma$~production in the leptonic $2\ell \gamma$~\cite{Aad:2019gpq} and $2\nu \gamma$ final states~\cite{Aaboud:2018jst} to constrain the $\CP$-odd interactions induced by $h_1^V$ and $h_2^V$, compare Eq.~\eqref{eq:nTGC_para}.
The measured inclusive cross section for $2 \ell \gamma$ final state is $\sigma_{2\ell\gamma} = (1065.4 \pm 23.5) \fb$~\cite{Aad:2019gpq}. For the analysis of the $2\nu\gamma$ final state which vetoes additional jets, the measured cross section is $\sigma_{2\nu\gamma} = (52.4 \pm 4.8) \fb$~\cite{Aaboud:2018jst}. 
We assume a detector efficiency of $\epsilon^\text{det}_{2\ell\gamma} = 0.54$ for the $2\ell \gamma$ channel and $\epsilon^\text{det}_{2\nu\gamma} = 0.89$ for the $2\nu\gamma$ channel. NNLO QCD and NLO EW corrections are again included through bin-by-bin $k$~factors by rescaling to the predictions in Refs.~\cite{Aad:2019gpq,Aaboud:2018jst}.

To calculate and combine the limits from $Z\gamma$, we again add up $\chi^2$ for each bin in the $E_{T, \gamma}$~distribution, see Eq.~\eqref{eq:chi2_nTGC}, using the binning given in the corresponding experimental references excluding overflow bins. 
Our last bins, which have the greatest sensitivity to the nTGCs, range from  $E_{T,\gamma} \in [500, \, 1200]\gev$ for $2\ell \gamma$  and $E_{T,\gamma} \in [600, \, 1100]\gev$ for the $2\nu \gamma$ analysis. 
As we will point out below, including the overflow in the last bin has a severe impact on the limits on $h_2^V$. 
To validate our analysis, we have explicitly checked that we can reproduce the expected limits of the analysis of the $2\nu\gamma$~final state at a luminosity of $36.1 \ifb$~\cite{Aaboud:2018jst} when including the overflow in the last bin.

Combining the limits from the $2\ell\gamma$ and $2 \nu\gamma$ final states for a luminosity of $3000\ifb$, we find $95\,\%$~CL bounds of 
\begin{equation}
    |h_1^\gamma| < 2.7 \times  10^{-4} \, , \quad
    |h_1^Z| < 2.4 \times 10^{-4} \, , \quad
    |h_2^\gamma| < 6.1 \times  10^{-7} \, , \quad
    |h_2^Z| < 6.1 \times 10^{-7} \, , 
\end{equation}
for the $\CP$-odd nTGCs. These values assume the same relative systematic uncertainties as in the experimental references at $36.1\ifb$ and $139\ifb$.
\begin{table}[b]
    \centering
    \begin{tabular}{lcccc}
    \toprule
    luminosity [$\ifb$]     &  $|h_1^\gamma| \times 10^{4}$ & $|h_1^Z| \times 10^{4}$  &  $|h_2^\gamma| \times 10^7$ & $|h_2^Z| \times 10^7$\\
         \midrule
    $139$  & $3.6$ & $3.2$ & $8.1$ & $8.1$ \\
    $300$  & $3.2$ & $2.9$ & $7.3$ & $7.2$ \\
    $3000$  & $2.7$ & $2.4$ & $6.1$ & $6.1$ \\
     \midrule
    $300$ (half syst)  & $2.7$ & $2.4$ & $6.1$ & $6.1$ \\
    $3000$ (half syst)  & $2.0$ & $1.8$ & $4.5$ & $4.4$\\
    \bottomrule
    \end{tabular}
    \caption{Expected limits on nTGCs for the combination of the $Z\gamma \to 2 \ell\gamma$ and $Z\gamma \to 2 
    \nu \gamma$ analyses at different luminosities. 
    In the two bottom rows, we present the limits assuming that the relative systematic uncertainties in each bin have been halved with respect to the value quoted by the experimental collaborations.}
    \label{tab:limits_ZA_ana}
\end{table}
Including the overflow in the last bin, the limits on $h_1^V$ tighten by $\sim 20\,\%$. 
On the other hand, the limits on $h_2^V$ are much more severely affected; they are approximately halved when including the overflow in the last bin. 
This implies that care has to be taken when translating limits based on an analysis including the overflow bin such as Ref.~\cite{Aaboud:2018jst}  into, for instance, an EFT framework. 
We collect the limits for other luminosities in Tab.~\ref{tab:limits_ZA_ana}. 
Since for higher luminosities and a fixed binning the uncertainty on the last bin quickly becomes dominated by systematic effects, we also present limits assuming systematic uncertainties are reduced by a factor of two. 
Because the limits on $\CP$-even nTGCs are roughly equivalent to those on their $\CP$-odd counterparts we do not present them explicitly here.

\section{Conclusions and Outlook}
\label{sec:conclusions}

We have studied the constraints on $\CP$-odd anomalous triple gauge couplings from diboson production. 

For the TGCs involving $W$~bosons, we have analysed differential asymmetries in $\CP$-sensitive observables based on $\Delta \phi$ and present our results in the SMEFT framework at dimension-six. Marginalizing over the second Wilson coefficient, we can constrain the coefficients to $|c_{H\tilde{W}B}|/\Lambda^2 < 0.04 \tev^{-2}$ and $|c_{\tilde{W}}|/\Lambda^2 < 0.02 \tev^{-2}$ at $3000 \ifb$. 
The strongest limits stem from the analysis of $W\gamma$, $Wjj$ and $Zjj$ production. 
The improved limits on the coefficient $c_{H\tilde{W}B}$ with respect to limits resulting from Higgs observables, motivates a combination of Higgs, vector boson scattering and diboson data for a combined fit of $\CP$-violating operators. 

To constrain neutral triple gauge couplings, we combined the bounds from the leptonic decay channels of $ZZ$ and $Z\gamma$ production. 
The most severe limits are obtained from the high-$p_T$ regimes of differential distributions instead of $\CP$-sensitive observables due to vanishingly small SM--New Physics interference terms. 
The resulting combined limits on $\CP$-odd interactions at $3000\,\ifb$ are $|f_4^Z| < 7.2  \times 10^{-4}$, $|f_4^\gamma| < 6.1\times 10^{-4}$ from $ZZ$~production and 
$|h_1^\gamma| < 2.7 \times  10^{-4} $, $|h_1^Z| < 2.4 \times 10^{-4}$, $|h_2^\gamma|, \, |h_2^Z| < 6.1 \times  10^{-7}$
from $Z\gamma$~production. 
Limits on $h_2^V$ are significantly tighter when including the overflow above $\sim 1 \tev$ in the last bin. 
This should be taken into account when translating these limits to an EFT framework. 

In summary, we presented expected limits on $\CP$-odd anomalous triple gauge couplings for future runs of the LHC and thereby provided bounds on additional sources of $\CP$-violation in the SM. 

\section*{Acknowledgements}
PG, FK and MS are supported by the UK Science and Technology Facilities Council (STFC) under grant  ST/P001246/1. 
AB gratefully acknowledges support from the Alexander-von-Humboldt foundation as a Feodor-Lynen Fellow.
FK and MS are acknowledging support from the European Union's Horizon 2020 research and innovation programme as part of the Marie Sklodowska-Curie Innovative Training Network MCnetITN3 (grant agreement no. 722104).
MS is funded by the Royal Society through a University Research Fellowship (URF\textbackslash{}R1\textbackslash{}180549), and FK gratefully acknowledges support by the Wolfson Foundation and the Royal Society under award RSWF\textbackslash{}R1\textbackslash{}191029.

\FloatBarrier
\bibliographystyle{amsunsrt_modpp}
\bibliography{literature}

\end{document}